\documentclass[11pt]{article} 

\usepackage{helvet}

\newcommand*{\myEXPfont}{\fontfamily{cmr}\selectfont}

\usepackage{graphicx} 
\usepackage{mathtools}
\usepackage{amsmath}
\usepackage{multirow}
\usepackage[export]{adjustbox}
\usepackage[left=1in, right=1in, top=1in, bottom=1in]{geometry} 
\usepackage{booktabs}
\usepackage{rotating}
\usepackage{rotfloat}
\usepackage{float}
\usepackage{subfig}
\usepackage{url}
\usepackage{pdfpages}
\usepackage{rotating}
\usepackage[doublespacing]{setspace} 
\usepackage{csvsimple}
\usepackage{array}
\usepackage{wrapfig}
\usepackage[utf8]{inputenc}

\usepackage{cite}
\usepackage{tabularx}
\usepackage{float}

\usepackage{amsthm}                
\usepackage{amssymb}

\usepackage{multirow}

\usepackage{caption}

\usepackage{amsfonts}
\usepackage{amsmath} 
\usepackage{tikz}
\usetikzlibrary{positioning,shapes.geometric}

\usepackage{xcolor}

\usepackage{setspace}

\DeclareMathOperator{\E}{\mbox{{\myEXPfont E}}}

\usepackage{mathrsfs}

\makeatletter
\newcommand*{\indep}{%
  \mathbin{%
    \mathpalette{\@indep}{}%
  }%
}
\newcommand*{\nindep}{%
  \mathbin{
    \mathpalette{\@indep}{\not}
  }%
}
\newcommand*{\@indep}[2]{%
  \sbox0{$#1\perp\m@th$}
  \sbox2{$#1=$}
  \sbox4{$#1\vcenter{}$}
  \rlap{\copy0}
  \dimen@=\dimexpr\ht2-\ht4-.2pt\relax
  \kern\dimen@
  {#2}%
  \kern\dimen@
  \copy0 
} 
\makeatother

\usepackage{xr-hyper}

\definecolor{forestgreen}{RGB}{34,139,34}
\usepackage{hyperref}
\hypersetup{
    colorlinks=true,
    linkcolor=magenta,
    filecolor=magenta, 
    citecolor=forestgreen,      
    urlcolor=blue
}

\usepackage{sectsty}
\sectionfont{\large}

\usepackage{layout}

\usepackage{geometry}

\newcolumntype{C}[1]{>{\centering\arraybackslash}p{#1}}

\usepackage{authblk}

\usepackage{afterpage}
\usepackage{lscape}
\usepackage{graphicx}

\usepackage{datetime}
\def\paperversionmajor{5}
\def\paperversionminor{0}

\makeatletter
\newcommand*{\addFileDependency}[1]{
  \typeout{(#1)}
  \@addtofilelist{#1}
  \IfFileExists{#1}{}{\typeout{No file #1.}}
}
\makeatother

\newcommand*{\myexternaldocument}[1]{%
    \externaldocument{#1}%
    \addFileDependency{#1.tex}%
    \addFileDependency{#1.aux}%
}

\myexternaldocument{appendix}

\begin{document}

\title{Regression-based estimation of heterogeneous treatment effects when extending inferences from a randomized trial to a target population}

\author[1,2]{Sarah E. Robertson}
\author[3]{Jon A. Steingrimsson}
\author[2,4]{Issa J. Dahabreh}

\affil[1]{Department of Health Services, Policy \& Practice, Brown University School of Public Health, Providence, RI}
\affil[2]{CAUSALab, Harvard T.H. Chan School of Public Health, Boston, MA}
\affil[3]{Department of Biostatistics, Brown University School of Public Health, Providence, RI }
\affil[4]{Departments of Epidemiology and Biostatistics, Harvard T.H. Chan School of Public Health, Boston, MA}

\maketitle{}
\thispagestyle{empty} 

\newpage 
\thispagestyle{empty} 

\vspace{0.1in}
\noindent \textbf{Address for correspondence:} Dr. Issa J. Dahabreh, Department of Epidemiology, Harvard T.H. Chan School of Public Health, Boston, MA 02115; email: \texttt{idahabreh@hsph.harvard.edu}; phone: +1 (617) 495‑1000.

\vspace{0.1in}
\noindent \textbf{Type of manuscript:} Original research article.

\vspace{0.1in}
\noindent \textbf{Running head:} Regression-based estimation of heterogeneous treatment effects.

\vspace{0.1in}
\noindent \textbf{Conflict of interest:} None.

\vspace{0.1in}
\noindent \textbf{Sources of financial support:} This work was supported in part by Patient-Centered Outcomes Research Institute (PCORI) award ME-1502-27794 (Dahabreh).  

\vspace{0.1in}
\noindent \textbf{Data and computing code availability:} The data analyses in our paper used CASS research materials obtained from the National Heart, Lung, and Blood Institute (NHLBI) Biologic Specimen and Data Repository Information Coordinating Center. We have provided code for implementing the two-step estimation procedure to reproduce similar figures on a simulated dataset for the $\texttt{R}$ environment on GitHub (link omitted for review). 

\vspace{0.1in}
\noindent
\textbf{Word count:} abstract = 211; main text $\approx$ 4000.

\vspace{0.1in}
\noindent
\textbf{Abbreviations:} CASS = Coronary Artery Surgery Study; CATE = conditional average treatment effect; MI = myocardial infarction.

\newpage
\thispagestyle{empty} 


\vspace*{1in}
{\LARGE \centering Regression-based estimation of heterogeneous treatment effects when extending inferences from a randomized trial to a target population \par }

\vspace{1in}
\noindent
\textbf{Running head:} Regression-based estimation of heterogeneous treatment effects.

\vspace{0.3in}
\noindent
\textbf{Type of manuscript:} Original Research Article.

\vspace{0.3in}
\noindent
\textbf{Word count:} abstract = 211; main text $\approx$ 4000.

\vspace{0.3in}
\noindent
\textbf{Abbreviations that appear in the text:} CASS = Coronary Artery Surgery Study. CATE = conditional average treatment effect. MI = myocardial infarction.

\clearpage
 \pagenumbering{arabic}
\setcounter{page}{1}
\vspace*{0.75in}
\begin{abstract}
\noindent
\linespread{1.3}\selectfont
Most work on extending (generalizing or transporting) inferences from a randomized trial to a target population has focused on estimating average treatment effects (i.e., averaged over the target population's covariate distribution). Yet, in the presence of strong effect modification by baseline covariates, the average treatment effect in the target population may be less relevant for guiding treatment decisions. Instead, the conditional average treatment effect (CATE) as a function of key effect modifiers may be a more useful estimand. Recent work on estimating target population CATEs using baseline covariate, treatment, and outcome data from the trial and covariate data from the target population only allows for the examination of heterogeneity over distinct subgroups. Here, we describe flexible pseudo-outcome regression modeling methods for estimating target population CATEs conditional on arbitrary discrete or continuous baseline covariates when the trial is embedded in a sample from the target population (i.e., nested trial designs). We construct pointwise confidence intervals for the CATE at specific values of the effect modifiers and uniform confidence bands for the CATE function. Last, we illustrate the methods using data from the Coronary Artery Surgery Study (CASS) to estimate CATEs conditional on history of previous myocardial infarction and baseline ejection fraction value in the target population of all trial-eligible individuals. 
\end{abstract}


\clearpage

\section*{INTRODUCTION}

When treatment effect modifiers have a different distribution among participants in a trial compared to the target population of substantive interest, the average treatment effect estimate from the trial is not directly applicable to the target population. A growing literature describes methods for extending -- transporting or generalizing \cite{hernan2016discussionkeiding, dahabreh2019commentaryonweiss} -- inferences about the average treatment effect from the trial to the target  \cite{cole2010, westreich2017, rudolph2017, dahabreh2018generalizing, dahabreh2020transportingStatMed}. These methods critically depend on adjusting for a large number of baseline covariates to ensure that the randomized trial and target population are conditionally exchangeable, allowing estimation of the target population average treatment effect.

Yet, in the presence of effect modification, the target population average treatment effect may not be sufficient for guiding treatment or policy decisions \cite{dahabreh2016}. Instead, the conditional average treatment effect (CATE) as a function of key effect modifiers may be a more useful estimand \cite{seamans2021generalizability}. Recent work \cite{mehrotra2021transporting} has described methods for estimating CATEs over distinct subgroups, when extending inferences from a trial to the target population. These methods, however, cannot be used to examine heterogeneity over continuous covariates or over multiple discrete covariates.

Here, we build on recent advances in estimating CATEs in observational studies \cite{abrevaya2015estimating, lee2017doubly, lechner2018modified, kennedy2020optimal, semenova2020debiased, fan2020estimation, knaus2021machine} and efficient and robust methods for generalizability and transportability analyses \cite{dahabreh2018generalizing, dahabreh2020transportingStatMed} to propose general methods for estimating the target population CATEs when a trial is embedded in a sample from the target population (i.e., in nested trial designs \cite{dahabreh2020studydesigns}). The methods allow the examination of CATEs given a small set of effect modifiers, while conditioning on a rich set of covariates that is sufficient to render the trial and target population exchangeable. Our approach involves two steps: In the first step, we create pseudo-outcomes using models for the probability of trial participation and the expectation of the outcome, possibly estimated using data-adaptive approaches (e.g., machine learning methods), to account for differences between the trial and the target population. In the second step, we regress the estimated pseudo-outcomes on effect modifiers of substantive interest to estimate the CATE as a function of key effect modifiers. We construct pointwise confidence intervals for the CATE at specific values of the effect modifiers and uniform confidence bands for the CATE function. We illustrate the methods using data from the Coronary Artery Surgery Study (CASS) to estimate CATEs conditional on history of previous myocardial infarction and baseline ejection fraction value in the target population of all trial-eligible individuals.

\section*{STUDY DESIGN, DATA, AND CAUSAL ESTIMANDS} \label{section_contrast}

\paragraph{Study design:} We consider a nested trial design \cite{dahabreh2020studydesigns}, where the trial is embedded within a cohort sampled from the target population of substantive interest. The nesting can be achieved by designing a prospective cohort study of individuals from the target population and inviting some of the cohort members to participate in the trial, while collecting information on baseline covariates on all cohort members, including those who do not participate in the trial. Nesting can also be achieved by retrospectively linking records from a completed trial with records from a cohort that is sampled from the target population. Regardless of how nesting is achieved, we assume that the cohort in which the trial is nested can be viewed as a simple random sample from the target population \cite{robins1988confidence}. Nested trial designs can be used for generalizability analyses, when the target population represented by the cohort meets the trial eligibility criteria, as well as for transportability analyses, when the target population represented by the cohort is broader than the population defined by the trial's eligibility criteria (see \cite{dahabreh2019commentaryonweiss, hernan2016discussionkeiding} for details regarding the definitions of the terms generalizability and transportability that we use in this paper; these terms are not used consistently in the literature, e.g., reference \cite{westreich2017} suggests different definitions).

\paragraph{Data:} In the nested trial design, data on a vector of baseline covariates, $X$, are available from all individuals in the cohort, regardless of participation in the trial. Data on the assigned treatment, $A$, and the outcome, $Y$, need only be available among trial participants. We use $S$ as the indicator for trial participation ($S = 1 $ for randomized individuals; $S=0$ for non-randomized individuals). Thus, the observed data are realizations of independent random draws of the tuple $O_i = (X_i, S_i, S_{i}A_{i}, S_{i}Y_{i})$, for $i, \ldots, n$, where $n$ is the total number of individuals in the cohort (both randomized and non-randomized). For simplicity, we assume that treatment $A$ is binary; extensions to address multi-valued treatments are possible. Throughout, italic upper-case letters denote random variables and corresponding lower-case letters denote realizations. We use $f(\cdot)$ to generically denote densities. 

\paragraph{Causal estimands:} Let $Y^a$ denote the potential (counterfactual) outcome under each treatment $a \in \{0,1\}$ \cite{splawaneyman1990, rubin1974, Robins1986}, that is, the outcome that would be observed under intervention to set treatment to $a$. Furthermore, let $\widetilde X$ denote a vector that contains a subset of the covariates in $X$ that, on the basis of mechanistic understanding and prior empirical evidence, are \emph{a priori} considered as the ``key'' effect modifiers under consideration. We are interested in the target population CATE given $\widetilde X = \widetilde x$, $$\E \big [ Y^1 - Y^0 \big | \widetilde X = \widetilde x \big ] = \E \big [ Y^1 \big | \widetilde X = \widetilde x \big ] - \E \big [ Y^0 \big | \widetilde X = \widetilde x \big ].$$ The CATE at some specific effect modifier value $\widetilde x$ is sometimes referred to as the ``group-average treatment effect'' \cite{lechner2018modified}.

In generalizability and transportability analyses, the covariate vector $X$ may be very high-dimensional because investigators collect information on multiple covariates in the hope of rendering the randomized trial and the target population exchangeable (in the sense formalized in the next section). In contrast, the vector of key effect modifiers $\widetilde X$ is typically of lower dimensionality as it only contains the small subset of baseline covariates that are deemed to be key effect modifiers. For instance, in our illustrative example based on the CASS, the investigators identified history of previous myocardial infarction and abnormal left ventricular function (defined as ejection fraction $<$50\%) as potentially important effect modifiers and examined them in subgroup analyses using data from trial participants \cite{investigators1984, passamani1985randomized}. Thus, we may want to examine whether these key covariates are also effect modifiers in the target population. Such examination, however, is likely to require conditioning on many additional covariates to render the trial participants exchangeable with the target population.


\section*{IDENTIFICATION} \label{section_identifiability}

\paragraph{Identifiability conditions:} The following conditions, which are sufficient for identifying the average treatment effect in the target population \cite{dahabreh2018generalizing, dahabreh2019identification}, are also sufficient for identifying CATEs in the target population:  

\noindent
\emph{(1) Consistency of potential outcomes:} if $A_i = a$, then $Y_i = Y^{a}_i$, for each $a \in \{0, 1\}$ and for each individual $i$ in the target population.

\noindent
\emph{(2) Mean exchangeability over $A$ in the trial:} for every $a \in \mathcal A$ and every $x$ with positive density in the trial $f(x, S = 1 ) \neq 0$, $\E [ Y^{a} | X = x , S = 1, A =a] = \E [ Y^{a} | X = x, S = 1]$.

\noindent
\emph{(3) Positivity of treatment assignment in the trial:} $\Pr[A=a | X = x, S=1] > 0$ for each $a$ and each $x$ with positive density in the trial $f(x , S = 1) \neq 0$.

\noindent
\emph{(4) Exchangeability in effect measure over $S$:} $\E [ Y^1 - Y^0 | X = x , S = 1] = \E [ Y^1 - Y^0 | X = x ] $, for every $x$ with positive density in the target population $f(x) \neq 0$.


\noindent
\emph{(5) Positivity of trial participation:} $\Pr[S=1 | X = x] >0,$ for every $x$ such that $f(x) \neq 0.$ 

The consistency condition over all individuals in the target population implicitly requires the absence of ``hidden'' versions of treatment \cite{rubin1986, rubin2010reflections, vanderWeele2009}, trial engagement effects \cite{dahabreh2019identification, dahabreh2019commentaryonweiss}, and interference \cite{rubin1986, halloran1995causal}. These assumptions are largely untestable and need to be considered on the basis of substantive knowledge. The conditions of mean exchangeability and positivity of treatment assignment are expected to hold in (marginally or conditionally) trials comparing well-defined interventions \cite{hernan2020causal}. The condition of exchangeability in effect measure over $S$ reflects an untestable assumption of a lack of effect measure modification by trial participation, conditional on baseline covariates, and needs to be examined in light of substantive knowledge and subjected to sensitivity analyses \cite{dahabreh2019sensitivitybiascor}. Last, positivity of trial participation is in principle testable, but its assessment can be challenging when $X$ is high-dimensional\cite{petersen2012diagnosing}. 

Of note, though we focus on generalizing inferences from a randomized trial, the methods we describe also apply to generalizing inferences from an observational study nested in a broader cohort, provided we are willing to assume that conditions (2) and (3) hold in observational study (i.e., no unmeasured baseline confounding and positivity of treatment within levels of the observed confounders) \cite{dahabreh2020benchmarking}.

\paragraph{Identification of CATEs:} In Web Appendix 1, we show that, under conditions (1) through (5), the target population CATE given $\widetilde X = \widetilde x$, that is, $\E \big [ Y^1 - Y^0 \big | \widetilde X = \widetilde x \big ]$, is identified by
\begin{equation} \label{identification_tedr}
  \delta( \widetilde x) \equiv   \E\big [ \phi(O) \big |  \widetilde X = \widetilde x \big];
\end{equation} 
where $$ \phi(O) = \dfrac{S (A -  e_1(X))}{ p(X)  e_1(X) e_0(X)} \big\{ Y -  g(X, A) \big\} +  g_1(X) -  g_0(X);$$
$p(X) = \Pr[S=1|X]$ is the conditional probability of trial participation given covariates; $e_a(X) = \Pr[A = a | X, S=1]$ is the conditional probability of treatment $a$ in the trial given covariates; $g_a(X) = \E[Y | X, S = 1, A = a]$ is the conditional expectation of the outcome in the trial given covariates, among individuals assigned to treatment $a \in \{0,1\}$; and $g(X, A) = \E[Y|X, S=1, A] = A g_{1}(X) + (1 - A) g_{0}(X)$. We refer to the functions $p(X)$, $e_a(X)$, and $g_{a}(X)$ as ``nuisance functions'' because they are useful in identifying and estimating CATEs, but, in our setup, are not of scientific interest \emph{per se}. Because $\phi(O)$ involves the observed data $O$ and nuisance functions that are identifiable from the observed data under the nested trial design \cite{dahabreh2020studydesigns}, we conclude that CATEs conditional on $\widetilde X$ are identifiable. In fact, $\phi(O)$ is the (uncentered) influence function of the functional that identifies the average treatment effect in the target population under a nonparametric model for the observed data that obeys conditions (1) through (5); see reference \cite{dahabreh2018generalizing} for details. Note also that setting the $g_{1}(X)$, $g_{0}(X)$, and $g(X, A)$ terms to zero in the expression for $\phi(O)$ gives an inverse probability weighting (IPW) identification result (see Web Appendix 1).

\section*{ESTIMATION \& INFERENCE}

\subsection*{Two-step estimation procedure} 

To extend causal inferences about CATEs given $\widetilde X$ from the trial to the target population, we propose to use a two-step procedure, similar to previously described approaches for estimating CATEs in observational analyses with baseline confounding by measured variables \cite{semenova2020debiased, fan2020estimation}. In the first step, we create pseudo-outcomes using models for the probability of trial participation, the expectation of the outcome, and (optionally) the probability of treatment in the trial, to account for differences between the trial and the target population. In the second step, we regress the estimated pseudo-outcomes on key effect modifiers to estimate the CATE function. It is important to make reasonable modeling choices in both steps, to learn about target population CATEs.

\paragraph{Step 1: estimation of the nuisance functions to calculate pseudo-outcomes:} We begin by estimating the nuisance functions, $p(X)$, $g_{a}(X)$, and $e_a(X)$. The estimated nuisance functions and the data are then used to calculate a pseudo-outcome for each observation: 
\begin{equation}\label{eq:estimated_pseudo}
\widehat \phi(O_i) = \dfrac{S_i (A_i - \widehat e_1(X_i))}{\widehat p(X_i) \widehat e_1(X_i) \widehat e_0(X_i)} \big\{ Y_i - \widehat g(X_i, A_i) \big\} + \widehat g_1(X_i) - \widehat g_0(X_i),
\end{equation}
where hats denote estimators of the corresponding nuisance functions in $\phi(O)$ and $\widehat g(X, A) = A  \widehat g_1(X) + (1 - A) \widehat g_0(X)$.

There are several options for estimating the nuisance functions $p(X)$, $g_{a}(X)$, and $e_a(X)$. The probability of treatment in the trial, $e_a(X)$, is known by design, and thus its estimation is straightforward using simple parametric models (e.g., logistic regression) \cite{williamson2014variance, lunceford2004}. In contrast, the functions $p(X)$ and $g_{a}(X)$ are unknown and involve conditioning on the high-dimensional baseline covariates that are necessary for exchangeability to hold between the trial and the target population. Thus, parametric models for the probability of participation and the expectation of the outcomes are likely to be misspecified, potentially leading to biased estimation of the CATE function. The pseudo-outcomes in equation \eqref{eq:estimated_pseudo} are chosen to render the procedure model doubly robust \cite{bang2005}, in the sense that the CATE function can be consistently estimated when at least one of these parametric models is correctly specified. Nevertheless, parametric models may poorly approximate both $p(X)$ and $g_{a}(X)$, in which case more flexible modeling strategies can help mitigate bias due to misspecification. The high-dimensionality of $X$ precludes fully nonparametric estimation \cite{robins1997toward}; instead, we can use data-adaptive modeling approaches to mitigate misspecification and allow more flexible modeling of the nuisance functions. Data adaptive approaches converge to the true underlying nuisance function at a slower rate than parametric models. To avoid bias when using data adaptive approaches, we rely on the estimators of the nuisance functions to closely approximate both nuisance functions for $p(X)$ and $g_{a}(X)$ at ``fast enough'' combined rate (i.e., the procedure has a rate-robustness property \cite{smucler2019unifying}). Many data-adaptive methods have rates that are fast enough (e.g., the highly adaptive lasso (HAL) \cite{benkeser2016highly} and generalized additive models (GAMs) \cite{horowitz2009semiparametric, kennedy2019robust}).

Background knowledge about aspects of the data-generating process can be used to select approaches that produce good approximations of the nuisance functions. For example, if we expect the relationship between trial participation status and covariates, or the outcome and covariates, to be highly nonlinear or to involve statistical interactions among predictors, random forest methods \cite{breiman2001random} may be a good choice to estimate the nuisance functions. If we expect sparsity, the least absolute shrinkage and selection operator (LASSO) \cite{tibshirani1996regression} or other sparsity-appropriate modeling approaches may be preferred.

\paragraph{Step 2: pseudo-outcome regression:} We fit a regression of the pseudo-outcomes on the effect modifiers of interest $\widetilde X$, to estimate the target population CATE as a function of $\widetilde x$,
\begin{equation} \label{tedr}
   \widehat\delta( \widetilde x) = \widehat  \E\big[\widehat{\phi}(O) \big| \widetilde X = \widetilde x \big].
\end{equation}

To estimate the CATE function, we need to correctly specify the pseudo-outcome regression model in the second step. One approach is to use a parametric model (e.g., least squares regression) to model the relationship between the pseudo-outcomes and the effect modifiers $\widetilde X$, but correct model specification may be challenging if $\widetilde X$ contains continuous components. To mitigate the risk of model misspecification, given that in our setup $\widetilde X$ is low dimensional, it may be possible to use nonparametric regression to flexibly model the relationship between the pseudo-outcomes and the effect modifiers in $\widetilde X$. For example, when $\widetilde X$ contains discrete covariates, it is simple to split the data into subgroups defined by the different levels of the covariates and estimate the mean of the pseudo-outcomes within each subgroup -- a non-smooth nonparametric ``regression'' approach \cite{racine2008nonparametric}. When $\widetilde X$ also contains continuous components, we can use series \cite{semenova2020debiased} or kernel (local linear) regression \cite{fan2020estimation} methods. For example, if $\widetilde X$ consists of a single continuous covariate, we can use a series estimator in the second step by fitting an ordinary least squares regression of the pseudo-outcome on a polynomial of increasing degree (alternatively, we can use splines or other basis functions) \cite{semenova2020debiased}.

\subsection*{Inference} 

We now discuss how to obtain (1) pointwise confidence intervals for the estimated CATE at specific values of the effect modifiers of interest and (2) uniform confidence bands for the estimated CATE function. The inference strategies we describe are most appropriate when the CATE function in the second step of the two-step estimation procedure is estimated using least squares, series, or kernel regression, provided appropriate technical requirements are met (e.g., regularity conditions, undersmoothing in the kernel regression, etc.). A detailed examination of these technical requirements is outside the scope of our paper (e.g., see \cite{semenova2020debiased} for series regression and \cite{fan2020estimation} for kernel regression). Statistical inference when using approaches other than least squares, series, or kernel regression in the second step of the two-step estimation procedure would require case-by-case examination.

\paragraph{Pointwise inference:} Pointwise confidence intervals for the CATE at a specific $\widetilde x$ value can be obtained using standard approaches, provided certain technical conditions hold (e.g., see \cite{semenova2020debiased} for series regression and \cite{fan2020estimation} for kernel regression in the second step). Specifically, a $(1- \alpha)\%$ pointwise confidence interval at $\widetilde x$ is given by $(\widehat\delta(\widetilde x) \pm z_{1-\alpha/2} \times \widehat \sigma(\widetilde x)),$ where $z_{1-\alpha/2}$ is the $(1- \alpha/2)$ quantile of the normal distribution and $\widehat \sigma(\widetilde x)$ is the estimated standard error obtained using the nonparametric bootstrap \cite{efron1994introduction}. Alternatively, when using ordinary least squares or series regression in the second step of the CATE estimation procedure, the robust variance estimator (i.e., the Huber-White sandwich estimator) \cite{huber1967behavior, stefanski2002}, which is readily available in standard software packages, can be used to obtain $\widehat \sigma(\widetilde x)$.  

\paragraph{Uniform inference:} Uniform confidence bands are needed to obtain valid coverage over a set of $\widetilde X$ values that we wish to examine. Suppose, for concreteness, that we are using series regression in the second step of the CATE estimation procedure. Series regression involves an estimator of the CATE as a function of $\widetilde x$ with the form $\widehat\delta (\widetilde x) = m(\widetilde x)\widehat \beta$, where $m(\widetilde x)$ is a vector of series or sieve basis functions (e.g., polynomials, splines, or wavelets) of $\widetilde x$, and $\widehat \beta$ is the least squares estimator of the regression coefficients. We will evaluate $\widehat\delta (\widetilde x)$ over a set of grid points $\mathcal{P}$, where $\mathcal{P}$ is a subset of the possible values of the effect modifiers $\widetilde X$. We work with the grid points in $\mathcal{P}$ instead of all possible values of $\widetilde X$ to allow for the possibility that $\widetilde X$ has continuous components. One option is to use as grid points all the unique values of $\widetilde x$ observed in the data; another is to choose grid points that capture the ``interesting'' values of $\widetilde x$. To obtain uniform inference over the set $\mathcal{P}$, following \cite{semenova2020debiased, fan2020estimation}, we use a multiplier bootstrap procedure \cite{belloni2015some, belloni2018uniformly, van1996weak} (also known as wild or weighted bootstrap).

The multiplier bootstrap procedure requires repeatedly fitting a second step regression of the pseudo-outcomes on the effect modifiers of interest over many replicates (it does not, however, require re-estimating the nuisance functions in the first step). Specifically, in each replicate $b = 1, \ldots, B$, where $B$ is the total number of replicates, we draw a vector of independent random variables of length $n$ from a standard exponential distribution. We use the elements of the vector as weights to fit a new weighted second step series regression of the pseudo-outcomes on the effect modifiers of interest, resulting in a new set of coefficients $\widehat \beta^b$. We then use $\widehat \beta^b$ to calculate the fitted value $\widehat\delta^{b}_{}(\widetilde x)= m(\widetilde x)\widehat \beta^b$ for each grid point $\widetilde x \in \mathcal{P}$. Next, we calculate the bootstrapped t-statistic,  $$ \widehat t^{b}(\widetilde x) = \dfrac{\widehat\delta^{b}_{}(\widetilde x)- \widehat\delta_{}(\widetilde x)}{\widehat \sigma(\widetilde x)},$$ for each grid point $\widetilde x \in \mathcal{P}$; here, the estimated CATE at $\widetilde x \in \mathcal P$, $\widehat\delta_{}(\widetilde x)$ is obtained from the series regression using the original dataset, and its standard error, $\widehat \sigma(\widetilde x)$, is obtained as described for pointwise inference. Next, we calculate the maximum of the absolute value of the t-statistic over the grid points, $\widehat t_{max}^b=\sup_{\widetilde x \in \mathcal{P}} | \widehat t^{b}(\widetilde x)|$. Finally, we obtain a $(1- \alpha)\%$ uniform confidence band as $$ \big\{ \widehat\delta_{}(\widetilde x) \pm \widehat{C}_{1-\alpha} \times \widehat \sigma(\widetilde x) : \widetilde{x} \in \mathcal{P} \big\},$$ where $\widehat C_{1-\alpha}$ is the $1- \alpha$ quantile of the empirical distribution, over the $B$ replicates, of $\widehat t_{max}^b$.

\paragraph{Considerations when using data-adaptive approaches:} In Web Appendix 3, we discuss additional considerations when using data-adaptive approaches in the first step of the estimation procedure.

\section*{EXAMINING HETEROGENEITY IN CASS}

\subsection*{CASS design and data}

Coronary Artery Surgery Study (CASS) \cite{investigators1984, william1983} compared coronary artery bypass grafting surgery plus medical therapy (hereafter, ``surgery'') versus medical therapy alone in a randomized trial that was nested within a cohort of trial-eligible individuals with chronic coronary artery disease. Individuals were enrolled from August 1975 to May 1979 and followed-up for death up to December 1996. The cohort consisted of 2099 trial-eligible individuals of whom 780 participated in the trial.

The original CASS trial analysis prespecified variables that the investigators believed to be important effect modifiers and risk factors for the outcome of mortality \cite{investigators1984}. These variables included history of previous myocardial infarction and abnormal left ventricular function (defined as ejection fraction value less than 50\%). One analysis of the trial participants in CASS at 10-years of follow-up \cite{alderman1990ten} found no difference in survival probability between treatment groups in the overall trial sample, but found that patients with an ejection fraction less than 50\% had significantly improved survival with surgery (surgery 79\%, medical therapy 61\%). No other subgroup-specific benefits were found in the trial \cite{passamani1985randomized, alderman1990ten}. A re-analysis of both randomized and observational data from CASS found heterogeneity on the risk difference scale for mortality at 10-years of follow-up among subgroups defined by history of myocardial infarction and abnormal left ventricular function \cite{robertson2020assessing}. Furthermore, a meta-analysis of 7 early trials (including CASS) comparing surgery versus medical therapy found that abnormal left ventricular function was an important modifier for the effect of treatment on mean survival time and that patients patients with abnormal left ventricular function derived greater absolute benefit from surgery \cite{yusuf1994effect}. A more recent randomized trial reported that among patients with ischemic cardiomyopathy and low ejection fraction ($<$35\%), surgery was more beneficial than medical therapy \cite{velazquez2016coronary}. Thus, we decided to use the methods described above to explore whether history of myocardial infarction and ejection fraction (treated as a continuous variable) were indeed effect modifiers in the target population of all trial-eligible individuals.

A total of 1686 individuals had complete data on the baseline covariates we used in our analysis (731 randomized, 368 to surgery and 363 to medical therapy; 955 non-randomized, 430 receiving surgery and 525 medical therapy). Table \ref{cass_baseline} summarizes the basic descriptive statistics for the baseline covariates. In general, randomized and non-randomized individuals had similar characteristics, but non-randomized individuals were more likely to have taken a beta-blocker regularly, have a higher left main coronary percent obstruction, and have a higher left ventricular wall score.

\subsection*{Statistical methods}
We are interested in evaluating the target population CATEs for mortality at 10-years of follow-up, conditional on history of previous myocardial infarction and baseline ejection fraction value. We analyzed patients with a history of previous myocardial infarction ($n=986$) and without a history of previous myocardial infarction ($n=700$) separately, in both estimation steps to fit the nuisance functions and the pseudo-outcome regression. We obtained pointwise confidence intervals and uniform confidence bands (with $b=200$ replicates). 

In the first step to estimate the nuisance functions for the outcome and participation, we used parametric models (logistic regression) and included the main effects of all baseline covariates listed in Table \ref{cass_baseline}, except we modeled age and ejection fraction using B-splines (basis splines) \cite{racine2014primer} of order 3 (degree 2) with an interior knot placed at the median of age or ejection fraction. We modeled the outcome separately in each treatment arm, among the trial participants. Because the model for treatment in the trial cannot be misspecified, to estimate $e_{a}(X)$, we used a simple logistic model that included the main effects of age, severity of angina, ejection fraction value, systolic blood pressure, proximal left anterior artery percent obstruction, and left ventricular wall score \cite{dahabreh2018generalizing}.

In the second step, we fit the regression of the pseudo-outcome on ejection fraction. We used ordinary least squares with either a B-spline or polynomial of ejection fraction. We did not want to assume that the CATE function has a linear form, so we chose to use B-splines of order 3 with an interior knot placed at the median of ejection fraction. For the polynomial of ejection fraction, we set the degree to 3. These are flexible enough to capture any reasonable non linearity in the treatment effect over ejection fraction.

For comparison, we also estimated trial-only CATEs, by altering the above procedure and fitting the second step regression among trial-participants only and using the trial pseudo-outcome (which does not include the participation weight and is justified by randomization) \cite{kennedy2020optimal}. To evaluate the robustness of our results to model specification of the nuisance functions in the first step, we repeated our analyses using generalized additive models (GAMs) instead of parametric models. 

\subsection*{Results}

Figure 1 shows the estimated target population CATE function stratified by history of myocardial infarction over a range of ejection fraction values, when using parametric models in the first step and splines in the second step. The CATE functions for patients with and without a history of myocardial infarction show different patterns, but both look like they could be reasonably well-approximated by a linear fit. For patients with a history of myocardial infarction, the CATE function linearly increases from a risk difference of -0.25 for patients with an ejection fraction of 30 up to a risk difference of up to 0.25 for patients with an ejection fraction of 80. The uniform confidence band suggests that the data are incompatible with the hypothesis that the CATE function is constant at 0 (no effect), over the range of ejection fraction values we considered. In other words, among patients with a history of myocardial infarction, the treatment effect appears to vary over ejection fraction, suggesting that surgery may be more beneficial (compared to medical therapy) for patients with lower ejection fraction, compared to those with higher ejection fraction. 

For patients without a history of myocardial infarction, the CATE function linearly decreases from a risk difference of 0.25 for patients with an ejection fraction of 40 to slightly less than 0 for patients with an ejection fraction of 70; then it increases up to 0.25 for patients with an ejection fraction of 80. Because the uniform confidence band contains zero, across all levels of ejection fraction, the data are not incompatible with the hypothesis that the CATE function is constant over ejection fraction for patients without a history of myocardial infarction.

In Web Appendix 4, we provide additional results for the CASS analysis. Appendix Figure 1 shows that the second step fit with a polynomial instead of splines in the second step yields similar results. The trial-only CATE functions, provided in Appendix Figures 2 and 3 have a similar trend to the target population CATE functions. We also found that repeating the analysis with GAMs in the first step resulted in similar CATE functions; see Appendix Figures 4 through 
7. We have provided a simulated dataset and code for the $\texttt{R}$ environment \cite{2021currentRcitation} that implements the two-step estimation procedure and produces graphs of the CATE function on GitHub (link omitted for review). We provide additional details about the code in Web Appendix 5.

\section*{DISCUSSION}

In this paper, we propose a two-step pseudo-outcome regression procedure for estimating target population CATEs in nested trial designs used to extend inferences from a randomized trial to a target population. We also described how to obtain both pointwise confidence intervals for the CATE at specific effect modifier values and uniform confidence bands for the CATE function. This two-step procedure provides a regression framework for examining CATEs given discrete as well as continuous covariates, whereas previously proposed methods only allow the estimation of CATEs within subgroups defined by discrete covariates \cite{mehrotra2021transporting,robertson2021subgroup}. Even when all covariates of interest are discrete, working within a regression framework may be advantageous because it allows the representation of smoothness or homogeneity assumptions by omitting covariate-by-covariate product terms from the pseudo-outcome regression specification; such assumptions are not as easy to represent with previously proposed methods. 

We note the different roles of the baseline covariates in the two steps of the procedure: the first step ``controls'' for enough variables to address selective trial participation; the second step focuses on a much smaller set of key effect modifiers. This duality is analogous to the difference between the variables needed to address confounding and effect modifiers in previous work on estimating CATEs in observational studies with baseline confounding of the treatment -- outcome association \cite{abrevaya2015estimating,  lee2017doubly,lechner2018modified, zimmert2019nonparametric, kennedy2020optimal, fan2020estimation, semenova2020debiased,knaus2021machine}. In applications, examining heterogeneity over a lower dimensional set of covariates may be motivated by scientific or policy considerations. For example, key effect modifiers may be identified on the basis of previous investigations, in which case the methods described here can be used in confirmatory assessments of heterogeneity. Or, it may be desirable to base policy about covariate-dependent treatment decisions using only a subset of potential effect modifiers while excluding unacceptable ones (e.g., even if insurance status was an effect modifier we might prefer to average over it and examine heterogeneity over lab-measurements or past medical history, rather than use insurance status).  

It is possible to extend our methods to settings where $\widetilde X$ is of moderate to high dimensionality, or even substituting $X$ for $\widetilde X$, as may be the case in discovery-oriented investigations \cite{kunzel2019metalearners, nie2017quasi, athey2019estimating, kennedy2020optimal}. In such investigations, the study goal is to predict individualized responses for members of the target population and the second step of the procedure typically is modified to use data-adaptive approaches appropriate for high-dimensional predictors \cite{ kennedy2020optimal}. The development of methods for valid inference in this context are an area of active research. Broadly speaking, in the setting of high-dimensional predictors, there exists a trade-off between the flexibility of the model specification and the strength of the technical assumptions needed for valid estimation and inference \cite{chernozhukov2018generic}. 

\clearpage
\bibliographystyle{ieeetr}
\bibliography{bibliography_cate}

\clearpage
\section*{Table}

\begin{table}[H]
\caption{CASS baseline table (August 1975 to December 1996). $S=1$ indicates the randomized group; $S=0$ indicates the non-randomized group; $A=1$ indicates surgery; $A=0$ indicates medical therapy.}
\label{cass_baseline}
\centering
\small
\begin{tabular}{@{}lcccc@{}}
\toprule
 & $S=1, A=1$ & $S=1, A=0$ & $S=1$ & $S=0$ \\ \midrule
Number of patients  & 368 & 363 & 731 & 955 \\
Age & 51.42 (7.24) & 50.92 (7.41) & 51.17 (7.32) & 50.89 (7.73) \\
History of angina & 285 (77.4) & 282 (77.7) & 567 (77.6) & 760 (79.6) \\
Taken beta-blocker regularly & 163 (44.3) & 152 (41.9) & 315 (43.1) & 508 (53.2) \\
Taken diuretic regularly & 63 (17.1) & 50 (13.8) & 113 (15.5) & 145 (15.2) \\
Ejection fraction & 60.86 (13.04) & 59.83 (12.78) & 60.35 (12.91) & 60.16 (12.25) \\
Employed full-time & 264 (71.7) & 233 (64.2) & 497 (68.0) & 632 (66.2) \\  
Type of job &  & &  &  \\
\hspace{3mm} High physical labor job & 151 (41.0) & 142 (39.1) & 293 (40.1) & 340 (35.6) \\
\hspace{3mm} Low mental labor job & 129 (35.1) & 135 (37.2) & 264 (36.1) & 320 (33.5) \\
\hspace{3mm} High mental labor job & 88 (23.9) & 86 (23.7) & 174 (23.8) & 295 (30.9) \\  
Left ventricular wall score & 7.44 (2.89) & 7.30 (2.78) & 7.37 (2.84) & 7.07 (2.69) \\
Taken nitrates regularly & 205 (55.7) & 196 (54.0) & 401 (54.9) & 528 (55.3) \\
History of previous MI & 209 (56.8) & 228 (62.8) & 437 (59.8) & 549 (57.5) \\
Male & 35 (9.5) & 37 (10.2) & 72 (9.8) & 87 (9.1) \\  
Smoking status &  & &  &  \\
\hspace{3mm} Never smoked & 62 (16.8) & 54 (14.9) & 116 (15.9) & 157 (16.4) \\
\hspace{3mm} Former smoker & 164 (44.6) & 157 (43.3) & 321 (43.9) & 451 (47.2) \\
\hspace{3mm} Current smoker & 142 (38.6) & 152 (41.9) & 294 (40.2) & 347 (36.3) \\  
High limitation of activities & 165 (44.8) & 173 (47.7) & 338 (46.2) & 441 (46.2) \\
High recreational activity & 228 (62.0) & 219 (60.3) & 447 (61.1) & 616 (64.5) \\
Confirmed hypertension & 118 (32.1) & 108 (29.8) & 226 (30.9) & 260 (27.2) \\  
Diabetes status &  & &  &  \\
 \hspace{3mm} No diabetes & 325 (88.3) & 328 (90.4) & 653 (89.3) & 873 (91.4) \\
 \hspace{3mm} Uncertain diabetes & 13 (3.5) & 7 (1.9) & 20 (2.7) & 23 (2.4) \\
 \hspace{3mm} Confirmed diabetes & 30 (8.2) & 28 (7.7) & 58 (7.9) & 59 (6.2) \\ 
LMCA percent obstruction & 4.27 (11.87) & 2.78 (9.55) & 3.53 (10.80) & 5.76 (14.50) \\
PLMA percent obstruction & 36.44 (38.04) & 34.89 (36.95) & 35.67 (37.49) & 39.14 (38.73) \\
Any diseased proximal vessels & 222 (60.3) & 230 (63.4) & 452 (61.8) & 608 (63.7) \\
Systolic blood pressure & 130.28 (17.40) & 130.34 (18.72) & 130.31 (18.06) & 129.80 (18.23) \\ \bottomrule
\end{tabular}
\caption*{LMCA = left main coronary artery; MI = myocardial infarction; PLMA = proximal left anterior artery. For continuous variables we report the mean (standard deviation); for binary variables we report the number of individuals (percentage).}
\end{table}

\clearpage
\section*{Figures}

\begin{figure}[htp]
 \caption{Target population CATE function estimated using parametric models in the first step and spline regression in the second step.}
    \centering
      \includegraphics[width=14cm]{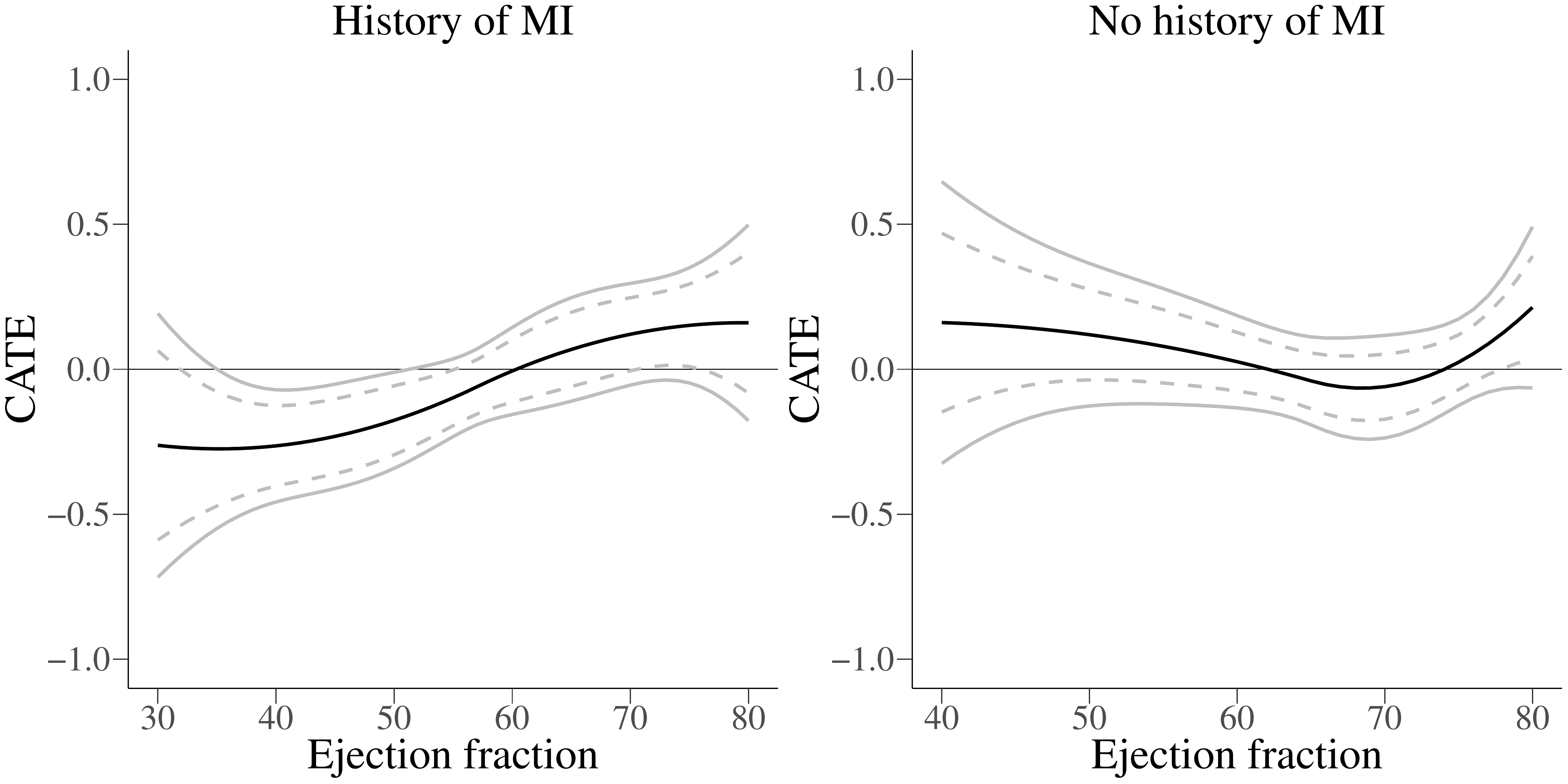}
     \caption*{ \\\
      \footnotesize CATE = conditional average treatment effect; MI = myocardial infarction. The black line indicates the estimated CATE function; dashed gray lines connect 95\% pointwise upper and lower confidence limits; solid gray lines depict the uniform 95\% confidence band. The set of grid points started at ejection fraction values of 30\% for patients with a history of myocardial infarction; 40\% for patients without a history of myocardial infarction because there are too few observations below 40\% for reliable interpretation. Both sets of grid points went up to ejection fraction values of 80\%, over a grid of evenly spaced points in steps of 1\%. 
      }
    \label{appendixfig:CATE_polynomial_all}
\end{figure}


\ddmmyyyydate 
\newtimeformat{24h60m60s}{\twodigit{\THEHOUR}.\twodigit{\THEMINUTE}.32}
\settimeformat{24h60m60s}
\begin{center}
\vspace{\fill}\ \newline
\textcolor{black}{{\tiny $ $generalizability\_HTE\_regression, $ $ }
{\tiny $ $Date: \today~~ \currenttime $ $ }
{\tiny $ $Revision: \paperversionmajor.\paperversionminor $ $ }}
\end{center}

\end{document}


\maketitle

\thispagestyle{empty}

\clearpage
\setcounter{tocdepth}{1}

\thispagestyle{empty}

\clearpage
\appendix 
\renewcommand{\thesection}{Web Appendix \arabic{section}}

\renewcommand{\thesubsection}{\arabic{section}.\arabic{subsection}}
\pagenumbering{arabic}

\renewcommand{\theequation}{\arabic{section}.\arabic{equation}}


\clearpage

\section{Identification}\label{appendix:identification}

Throughout this appendix we use the notation defined in the main text of the paper. 

\begin{proposition*}
If conditions (1) through (5) from the main text hold, then the conditional average treatment effect given $ \widetilde X = \widetilde x $, $ \emph{E} [ Y^1 - Y^0 | \widetilde X = \widetilde x ]$, is identified by $$   \delta( \widetilde x) \equiv   \text{\emph{E}} \big[ \phi(O) \big |  \widetilde X = \widetilde x \big],$$
where $$\phi(O) = \dfrac{S (A - e_1(X))}{p(X)e_1(X)e_0(X)} \big\{ Y - g(X, A) \big\} + g_1(X) - g_0(X).$$
\end{proposition*}

\vspace{0.1in}

\begin{proof}

First, we will show that, under conditions (1) through (5), \begin{equation}\label{eq:ipw_identification}
    \E[Y^1 - Y^0 | \widetilde X] = \E\left[ \dfrac{S (A - e_1(X))}{p(X)e_1(X)e_0(X)}Y \Big| \widetilde X \right].
\end{equation}
Starting from the right-hand-side,
\begin{equation*}
    \begin{split}
     & \E\left[ \dfrac{S (A - e_1(X))}{p(X)e_1(X)e_0(X)}Y \Big| \widetilde X \right]   \\
    &\quad\quad\quad= \E\Bigg[ \E\left[ \dfrac{ I(S = 1, A = 1)Y}{p(X) e_1(X)  } -  \dfrac{ I(S = 1, A = 0)Y}{p(X) e_0(X)  }  \Big |  X  \right] \Big| \widetilde X  \Bigg] \\
    &\quad\quad\quad= \E\Bigg[ \E\left[ \dfrac{ I(S = 1, A = 1)Y}{p(X) e_1(X)  }  \Big |  X  \right] - \E\left[ \dfrac{ I(S = 1, A = 0)Y}{p(X) e_0(X)  }  \Big |  X  \right]  \Big| \widetilde X  \Bigg] \\
    &\quad\quad\quad=\E \big[ \E[Y  | X, S = 1, A = 1] - \E[ Y | X, S = 1 , A = 0] \big| \widetilde X \big]  \\
            &\quad\quad\quad=\E \big[ \E[Y^1  | X, S = 1, A = 1] - \E[ Y^0 | X, S = 1 , A = 0] \big| \widetilde X \big]  \\
            &\quad\quad\quad=\E \big[ \E[Y^1  | X, S = 1] - \E[ Y^0 | X, S = 1 ] \big| \widetilde X \big]  \\
            &\quad\quad\quad=\E \big[ \E[Y^1-Y^0  | X, S = 1]  \big| \widetilde X \big] \\
          &\quad\quad\quad=\E \big[ \E[Y^1-Y^0  | X]  \big| \widetilde X \big] \\
           &\quad\quad\quad=\E \big[ Y^1-Y^0   \big| \widetilde X \big],
    \end{split}
\end{equation*}
where the first step follows from the law of total expectation; the second step by linearity of expectation; the third by the fact that $\E[I(S = 1, A = a) Y | X] = \E[Y | X, S = 1, A = a] \Pr[S = 1, A = a | X] = \E[Y | X, S = 1, A = a] p(X) e_a(X) $ for $a = 0, 1$ and cancellation of terms; the fourth by consistency; the fifth by conditional mean exchangeability over $A$; the sixth by linearity of expectation; the seventh by conditional mean exchangeability over $S$; and the last from the law of total expectation. All quantities are well-defined because of the positivity conditions.

Next, we will show that 
\begin{equation}\label{eq:if_identification}
    \E\big[\phi(O)  | \widetilde X \big] = \E\left[ \dfrac{S (A - e_1(X))}{p(X)e_1(X)e_0(X)}Y \Big| \widetilde X \right].
\end{equation}

Starting from the left-hand-side and using the definition of $\phi(O)$,
\begin{equation*}
    \begin{split}
        &\E\big[\phi(O) \big | \widetilde X \big] \\
            &\quad\quad\quad= \E\left[ \dfrac{S (A - e_1(X))}{p(X)e_1(X)e_0(X)} \big\{ Y - g(X,A) \big\} + g_1(X) - g_0(X) \Big | \widetilde X \right] \\
            &\quad\quad\quad= \E\left[ \dfrac{S (A - e_1(X))}{p(X)e_1(X)e_0(X)}Y \Big| \widetilde X \right]  -  \E\left[ \dfrac{S (A - e_1(X))}{p(X)e_1(X)e_0(X)}  g(X,A) \Big| \widetilde X \right] \\
            &\quad\quad\quad\quad\quad\quad\quad\quad\quad\quad\quad\quad + \E \big [g_1(X) - g_0(X)  \big| \widetilde X \big] \\ 
            &\quad\quad\quad= \E\left[ \dfrac{S (A - e_1(X))}{p(X)e_1(X)e_0(X)}Y \Big| \widetilde X \right],
    \end{split}
\end{equation*}
where the last step follows because $$ \E\left[ \dfrac{S (A - e_1(X))}{p(X)e_1(X)e_0(X)}  g(X, A) \Big | \widetilde X \right] = \E \big [g_1(X) - g_0(X)  \big| \widetilde X \big].$$ 

By combining the results in equations \eqref{eq:ipw_identification} and \eqref{eq:if_identification}, we can conclude that $$ \E[Y^1 - Y^0 | \widetilde X  = \widetilde x] =  \E\big[\phi(O) \big | \widetilde X  = \widetilde x \big] \equiv \delta( \widetilde x).$$ 

\end{proof}

\clearpage
\section{Estimation using inverse probability weighting}\label{appendix_estimation}

\paragraph{Estimating target population CATEs by modeling participation:} 

The sample analog of the identification result in equation \eqref{eq:ipw_identification} suggests the following two-step inverse probability weighting (IPW) approach for estimating the CATE function: 

In the first step, using the estimated nuisance functions for the probability of trial participation and the probability of treatment in the trial, we calculate inverse probability weighted pseudo-outcomes as $$\widehat \psi(O)= \dfrac{S (A - \widehat e_1(X))}{ \widehat p(X) \widehat e_1(X) \widehat e_0(X)}Y.$$ 

In the second step, we estimate the CATE function using a regression of the pseudo-outcomes on $\widetilde X$ to estimate 
\begin{equation*}
   \widehat\delta_{\text{\tiny IPW}}(\widetilde x) = \widehat \E \big[\widehat \psi(O) \big| \widetilde X = \widetilde x \big].
\end{equation*}

Related results for observational studies suggest that the estimator $\widehat\delta_{\text{\tiny IPW}}(\widetilde x)$ will generally be less efficient than the estimator $\widehat\delta(\widetilde x)$ described in the main text \cite{kennedy2020optimal}.

\clearpage
\section{Considerations when using data-adaptive approaches} \label{appendix:data_addaptive}
\renewcommand\thefigure{\arabic{figure}}   
\setcounter{figure}{0}
\setcounter{equation}{0}

In the first step of the estimation procedure, data-adaptive approaches for estimating the nuisance functions use regularization to trade off bias and variance. Heavy regularization can lower the rate of convergence of nuisance functions estimators of the  underlying true functions, resulting in regularization bias. The two-step estimation procedure mitigates regularization bias by combining models for both $p(X)$ and $g_{a}(X)$ (the rate robustness property \cite{smucler2019unifying} mentioned above). That is, the estimation procedure can still be used for valid inference, even if the estimators of the nuisance functions converge at a rate slower than the parametric rate, provided the ``combined'' rate of convergence is ``fast enough.'' A precise definition of what constitutes a fast enough combined rate of convergence is beyond the scope of this manuscript (interested readers may wish to consult references \cite{chernozhukov2018double, semenova2020debiased}). What is practically important, is that many popular machine learning methods do converge at a fast enough rate \cite{chernozhukov2018double}.

Furthermore, estimating the nuisance functions in the first step and the pseudo-outcome regression in the second step with the same data may induce correlations between estimation errors (i.e., the differences between the true functions and their estimators) and unobservable model errors, resulting in overfitting bias. Such bias may be particularly concerning when using highly data-adaptive methods, such as random forest or neural network methods, in the first step. To mitigate overfitting bias when implementing the two-step estimation procedure, it may be necessary to use cross-fitting \cite{chernozhukov2018double} to eliminate the correlation that induces overfitting bias. Note that however, that if the data-adaptive approach meets certain smoothness and regularity conditions \cite{semenova2020debiased}, as is the case for generalized additive models (GAMs) \cite{horowitz2009semiparametric} or the highly adaptive LASSO \cite{benkeser2016highly}, then cross-fitting may not be necessary. 

For simplicity, we shall describe cross-fitting with a single split into two separate folds \cite{semenova2020debiased}: (1) randomly split the data into two folds; (2) estimate the nuisance functions in the first fold and use the estimated functions to obtain the predictions in the second fold; (3) interchange the role of the two folds and repeat the previous step; (4) calculate the pseudo-outcomes and fit the pseudo-outcome regression in the full dataset. Cross-fitting variants using additional folds for different nuisance models, or estimating the pseudo-outcome regression within folds and then averaging across estimates from the different fitted regressions, have been described in the context of observational studies with baseline confounding by measured variables \cite{semenova2020debiased, fan2020estimation, kennedy2020optimal} and can be adapted to analyses extending inferences from a randomized trial to a target population.

\clearpage
\section{Additional results from the CASS analyses} \label{appendix:cass_baselines}
\renewcommand\thefigure{\arabic{figure}}   
\setcounter{figure}{0}
\setcounter{equation}{0}
\renewcommand{\figurename}{Appendix Figure}
\renewcommand{\tablename}{Appendix Table}

\paragraph{Target population CATE function estimated using parametric models in the first step and polynomial regression in the second step:} In the main text, Figure 1 shows the estimated target population CATE function stratified by history of myocardial infarction over a range of ejection fraction values, when using parametric models in the first step and splines in the second step. Here, we show the CATE function estimated with polynomial regression in the second step for comparison.
For all figures in this appendix we use the following abbreviations: CATE = conditional average treatment effect; MI = myocardial infarction. 

On the plots, black lines indicate the estimated CATE function; dashed gray lines connect 95\% pointwise upper and lower confidence limits; solid gray lines depict the uniform 95\% confidence band. The set of grid points started at ejection fraction values of 30\% for patients with a history of myocardial infarction; 40\% for patients without a history of myocardial infarction because there are too few observations below 40\% for reliable interpretation. Both sets of grid points went up to ejection fraction values of 80\%, over a grid of evenly spaced points in steps of 1\%.

\vspace*{0.2in}

\begin{figure}[htp]
 \caption{Target population CATE function estimated using parametric models in the first step and polynomial regression in the second step.}
    \centering
 \includegraphics[width=14cm]{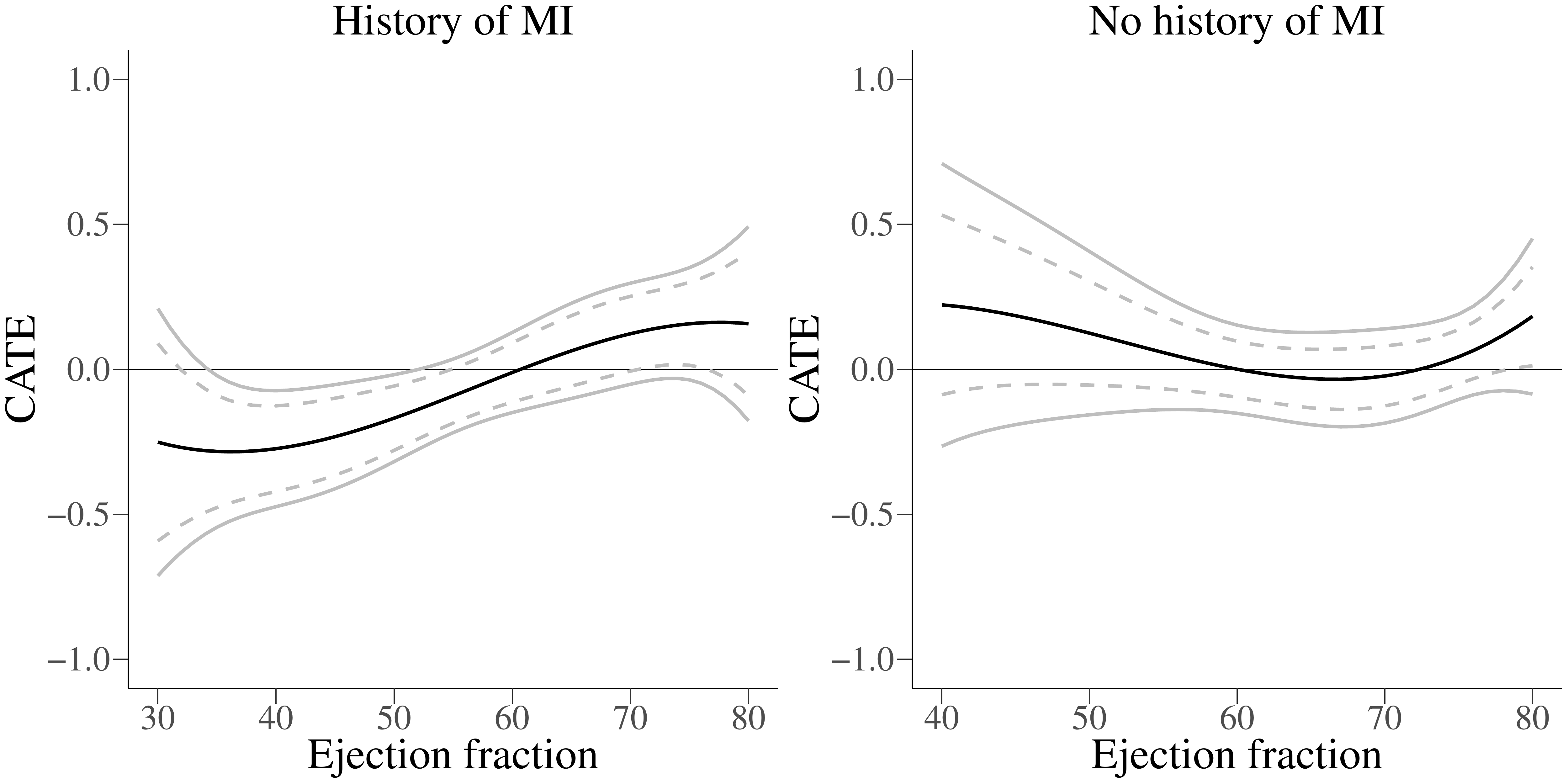}
      \label{appendixfig:CATE_splines_all}
\end{figure}

\clearpage
\paragraph{The CATE function in the trial:} Here, we plot the CATE function given history of previous myocardial infarction and baseline ejection fraction value, estimated using data only from trial participants.

\vspace*{0.2in}

\begin{figure}[htp!]
 \caption{Trial CATE function estimated using parametric models in the first step and polynomial regression in the second step.}
    \centering
    \includegraphics[width=14cm]{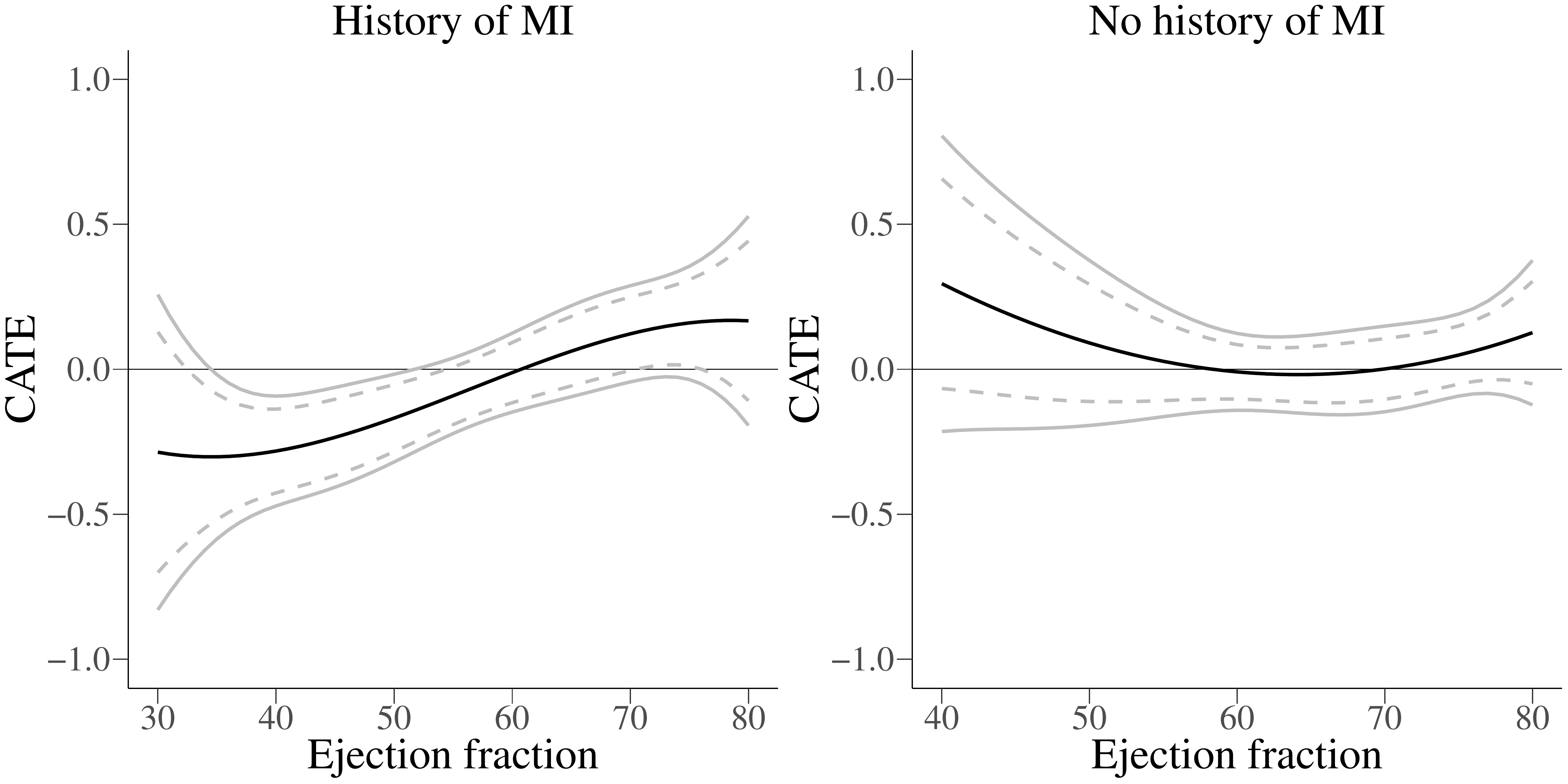}
    \label{appendixfig:CATE_polynomial_trial}
\end{figure}

\begin{figure}[htp!]
  \caption{Trial CATE function estimated using parametric models in the first step and spline regression in the second step.}
    \centering
    \includegraphics[width=14cm]{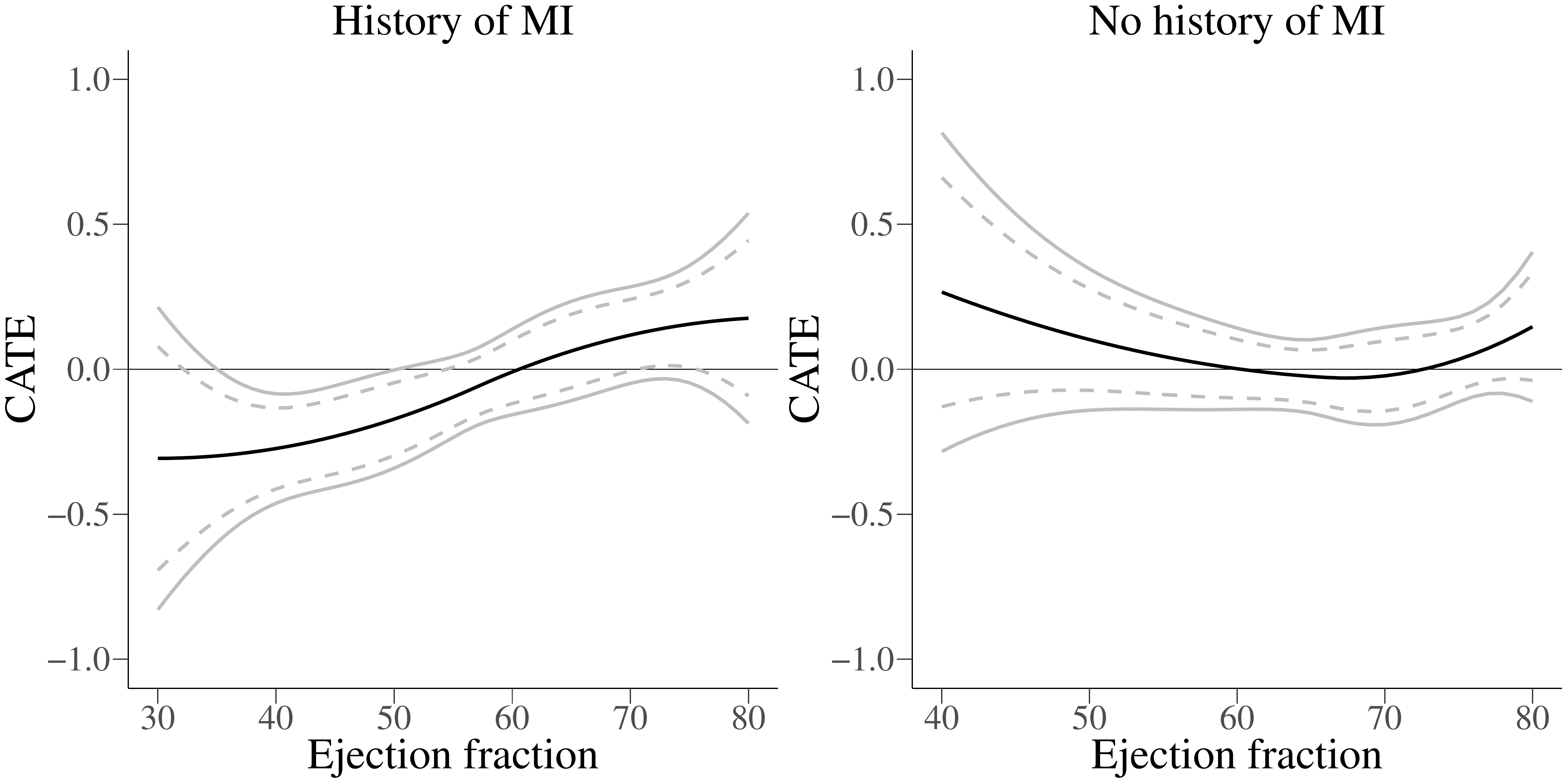}
    \label{appendixfig:CATE_splines_trial}
\end{figure}

\clearpage
\paragraph{Target population CATE function estimated using generalized additive models in the first step:} Here, we show the estimated target population and trial CATE functions using generalized additive models (GAMs) to estimate the nuisance functions in the first step, with a polynomial regression or splines in the second step. In the first step, we used a similar model specification as the one described in the main text. We used the \texttt{gam} function in the \texttt{mgcv} package (v1.8-34) \cite{Wood2011-ax, woodGAMbook} in \texttt{R} (v4.0.4) \cite{2021currentRcitation}, which fits a smoothness penalty using generalized cross-validation, and we considered splines for age and ejection fraction.  Because GAMs are ``smooth enough'', cross-fitting is not necessary in our application \cite{chernozhukov2018double}. In the second step, we used the model specification described in the main text.

\vspace*{0.2in}

\begin{figure}[htp]
 \caption{Target population CATE function estimated using generalized additive models in the first step and polynomial regression in the second step.}
    \centering
      \includegraphics[width=14cm]{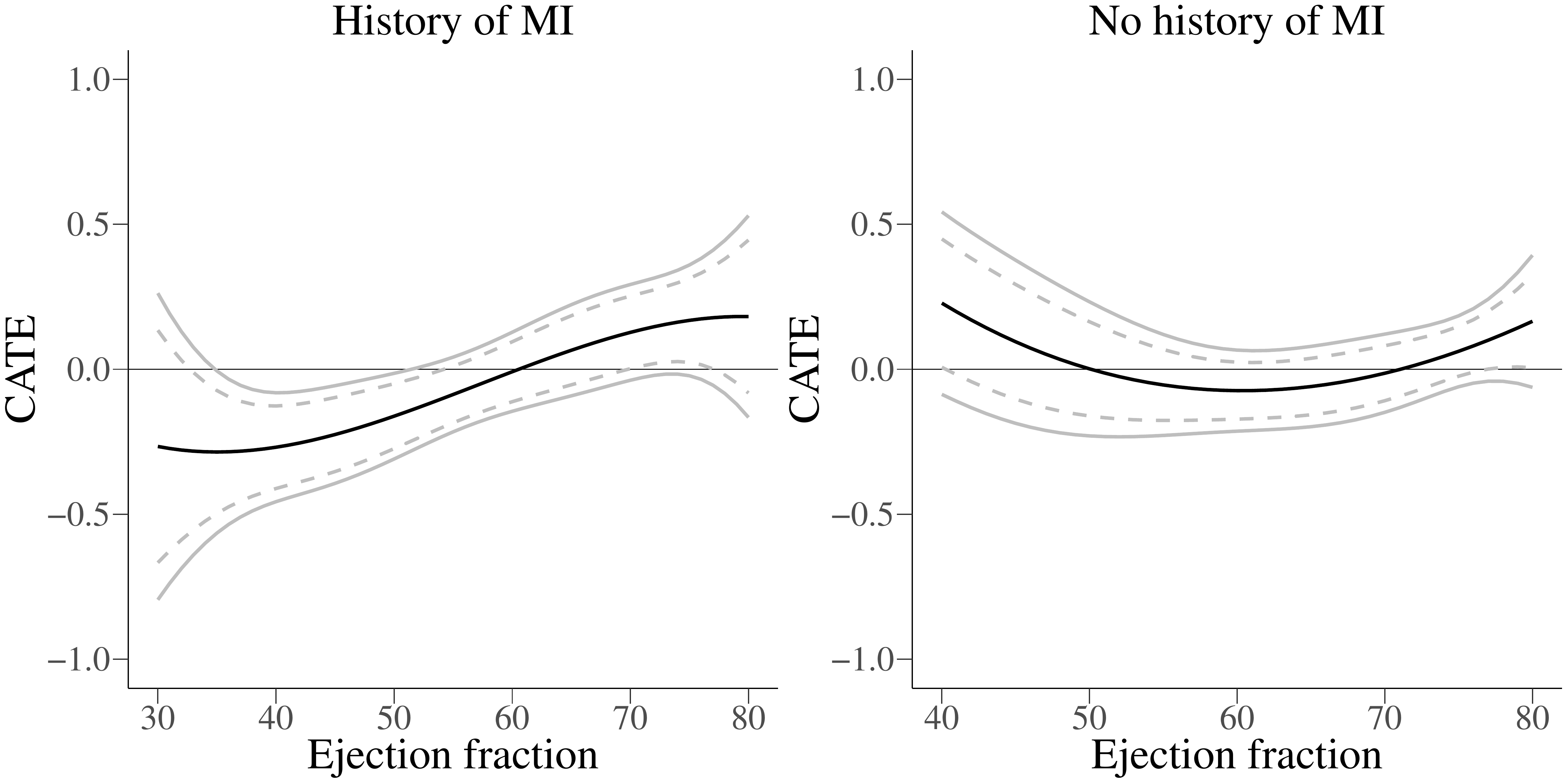}
    \label{fig:GAM_CATE_polynomial_all}
\end{figure}

\begin{figure}[htp]
 \caption{Target population CATE function estimated using generalized additive models in the first step and spline regression in the second step.}
    \centering
    \includegraphics[width=14cm]{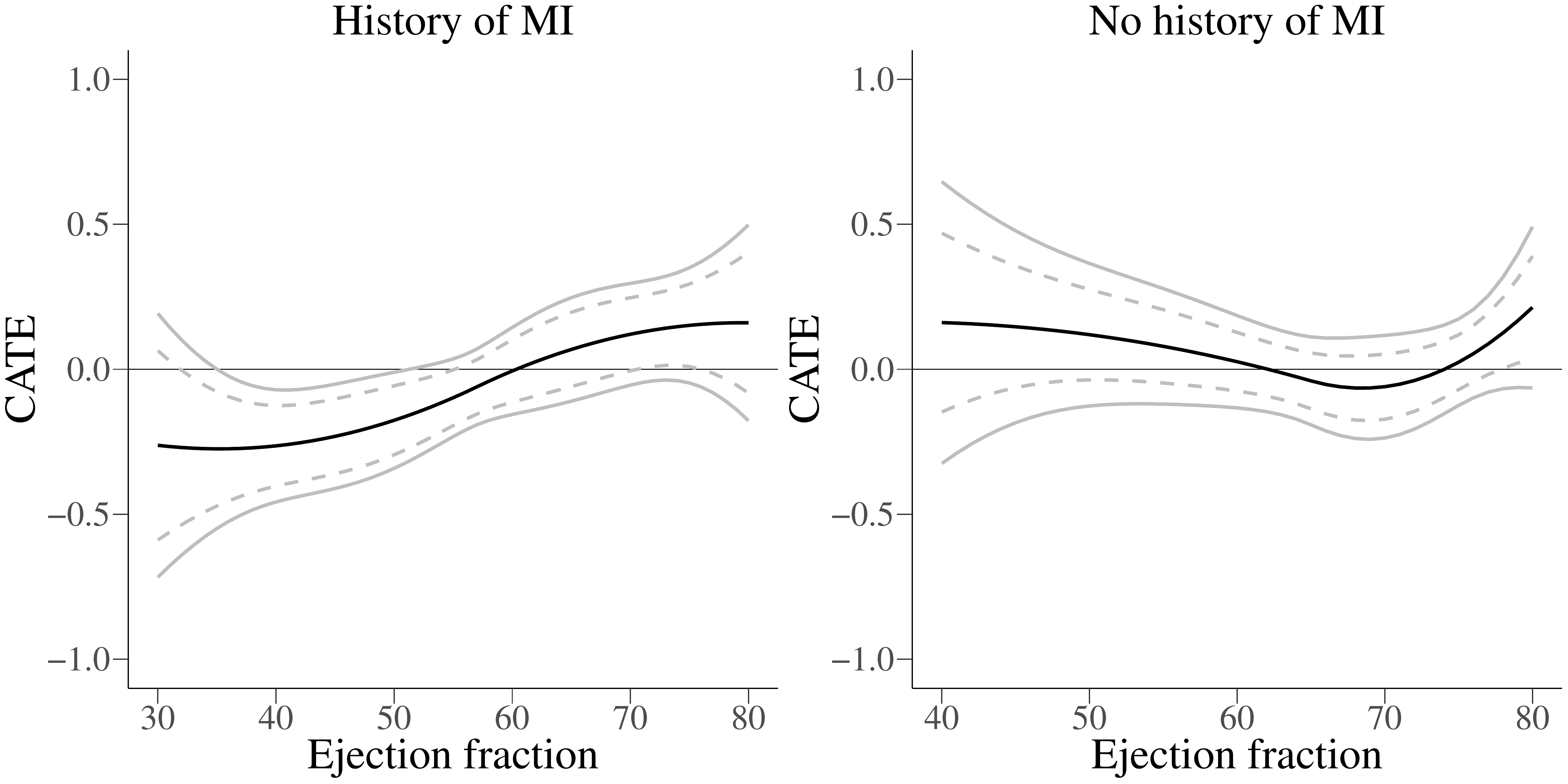}
    \label{fig:GAM_CATE_splines_all}
\end{figure}

\begin{figure}[htp]
 \caption{Trial CATE function estimated using generalized additive models in the first step and polynomial regression in the second step.}
    \centering
    \includegraphics[width=14cm]{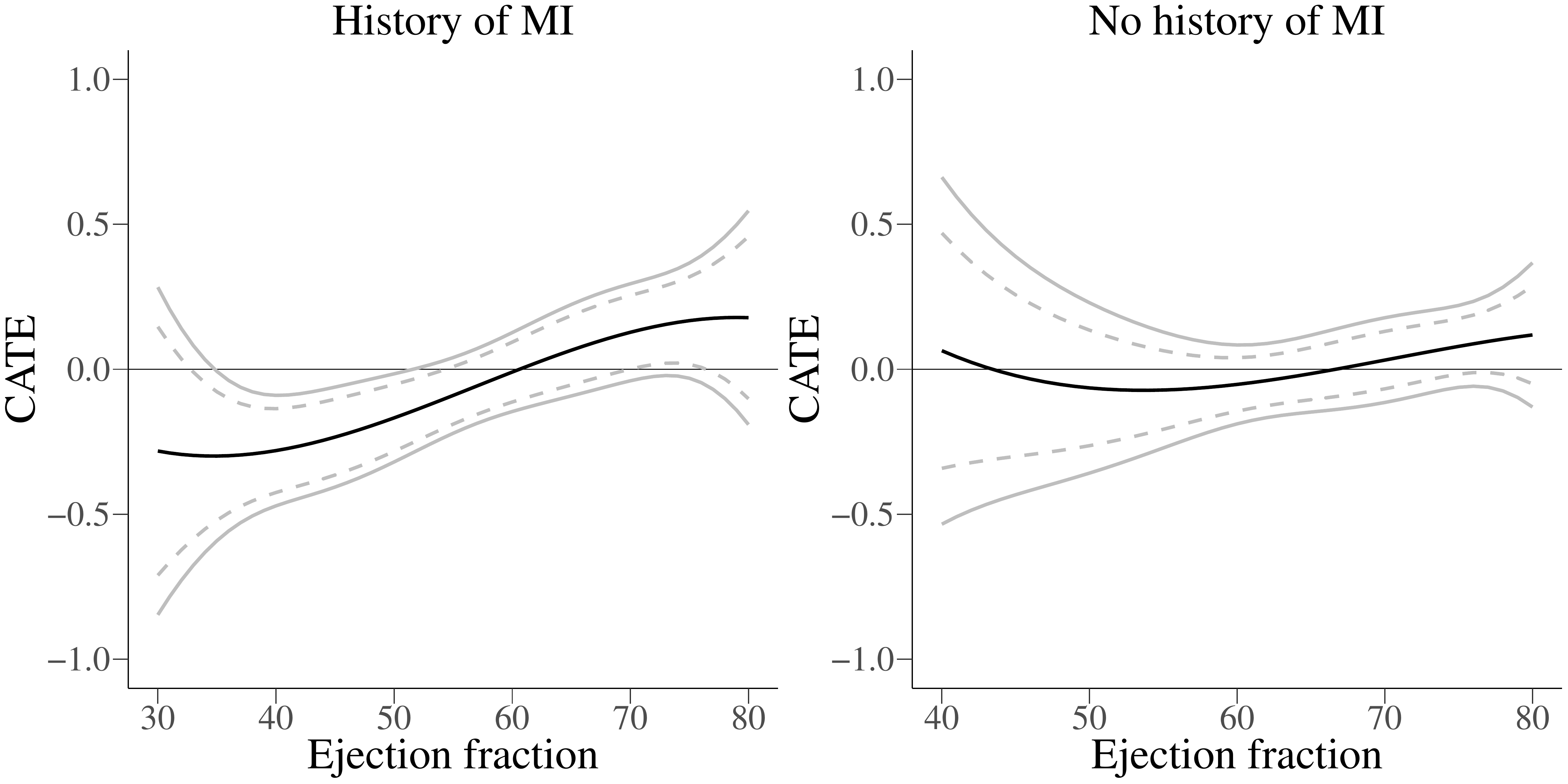}
    \label{appendixfig:GAM_CATE_polynomial_trial}
\end{figure}

\begin{figure}[htp]
 \caption{Trial CATE function estimated using parametric models in the first step and spline regression in the second step.}
    \centering
    \includegraphics[width=14cm]{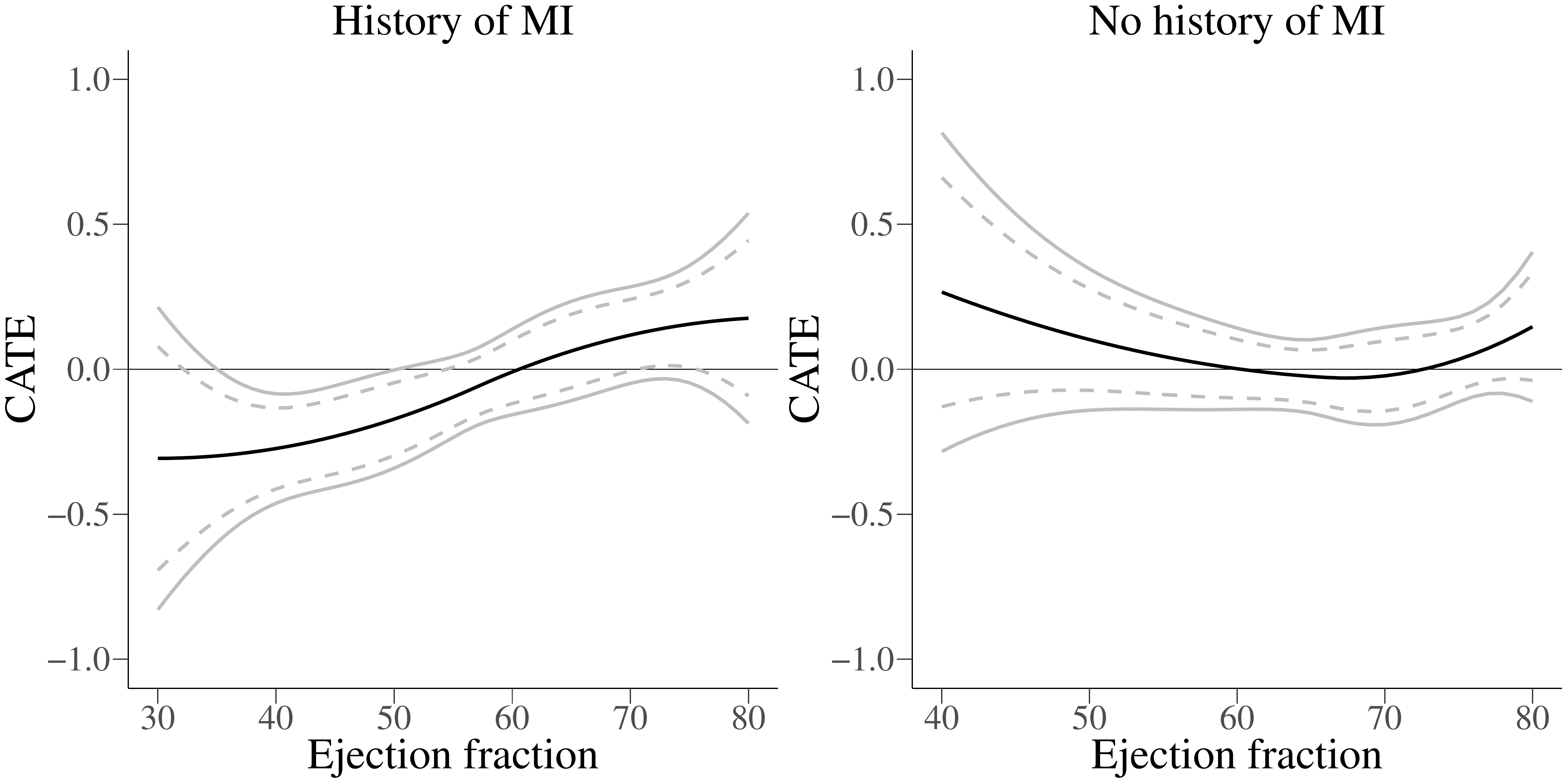}
    \label{appendixfig:GAM_CATE_splines_trial}
\end{figure}

\clearpage
\section{Code and data}\label{appendix:code}

\paragraph{Code for empirical analyses:}

We provide \texttt{R} code to implement the two-step estimation procedure described in the paper for the $\texttt{R}$ environment on GitHub on a simulated dataset for illustration: [link removed for peer review]. 

\paragraph{Data availability:} The CASS study data are not publicly available, but they can be obtained from the National Heart, Lung, and Blood Institute (NHLBI) Biologic Specimen and Data Repository Information Coordinating Center (\url{https://biolincc.nhlbi.nih.gov/studies/cass/}; last accessed April 30, 2021).

\clearpage
\bibliographystyle{unsrt}
\bibliography{bibliography_cate}{}